%% file: MACSJ0717_TSZX_map.tex
\documentclass[twocolumn,traditabstract]{aa}  

\usepackage{fixltx2e}
\usepackage[english]{babel}
\usepackage{graphicx,amsmath}
\usepackage{epstopdf}
\usepackage{epsf,color}
\usepackage[mathscr]{eucal}
\usepackage{amsmath}
\usepackage{amssymb,amsfonts}
\usepackage{natbib}
\usepackage{graphicx}
\usepackage{newtxtext}
\usepackage{microtype}
\usepackage{dsfont}
\definecolor{Mygreen}{rgb}{0.75, 0.0, 0.0}
\definecolor{Mypink}{rgb}{1.0, 0.0, 0.5}
\definecolor{Myred}{rgb}{0.7, 0.0, 0.0}
\usepackage[breaklinks, citecolor=blue, linkcolor=Myred, urlcolor=Myred, colorlinks=true, debug, baseurl=' ']{hyperref}
\usepackage{float} 
\usepackage{color}

\bibpunct{(}{)}{;}{a}{}{,}
\def\xe {n_{\rm e}}
\def\pe {P_{\rm e}}
\def\TSZ {T_{\rm SZX}}
\def\TMW {T_{\rm gmw}}
\def \TXC {T_{\rm CXO}}
\def \TXX {T_{\rm XMM}}
\def \TXXC {T_{\rm XMM/CXO}}
\def \TX {T_{\rm X}}

\begin{document}

\title{Mapping the hot gas temperature in galaxy clusters using X-ray and Sunyaev-Zel'dovich imaging}
\input{listeauthors}

\date{Received \today \ / Accepted --}
\abstract {We propose a method to map the temperature distribution of the hot gas in galaxy clusters that uses resolved images of the thermal Sunyaev-Zel'dovich (tSZ) effect in combination with X-ray data. Application to images from the New IRAM KIDs Array (NIKA) and XMM-\textit{Newton} allows us to measure and determine the spatial distribution of the gas temperature in the merging cluster \mbox{MACS~J0717.5+3745}, at $z=0.55$. Despite the complexity of the target object, we find a good morphological agreement between the temperature maps derived from X-ray spectroscopy only -- using XMM-\textit{Newton} ($T_{\rm XMM}$) and \textit{Chandra} ($T_{\rm CXO}$) -- and the new gas-mass-weighted tSZ+X-ray imaging method ($\TSZ$). We correlate the temperatures from tSZ+X-ray imaging and those from X-ray spectroscopy alone and find that $\TSZ$ is higher than $T_{\rm XMM}$ and lower than $T_{\rm CXO}$ by $\sim 10\%$ in both cases. Our results are limited by uncertainties in the geometry of the cluster gas, contamination from kinetic SZ ($\sim 10\%$), and the absolute calibration of the tSZ map ($7\%$). Investigation using a larger sample of clusters would help minimise these effects.}
\titlerunning{Mapping the temperature in galaxy clusters using X-ray and tSZ imaging}
\authorrunning{R. Adam, M. Arnaud, I. Bartalucci, et al.}
\keywords{Techniques: high angular resolution -- Galaxies: clusters: individual: \mbox{MACS~J0717.5+3745}; intracluster medium -- X-rays: galaxies: clusters}
\maketitle

\section{Introduction}\label{sec:Introduction}
In galaxy clusters, temperature and density are the key observable characteristics of the hot ionised gas in the intracluster medium (ICM). X-ray observations play a fundamental role in their measurement. Density is trivial to obtain from X-ray imaging, while temperature can be derived from an isothermal model fit to the spectrum. Accurate gas temperatures are needed for a number of reasons. Accurate temperatures are essential to infer cluster masses under the assumption of hydrostatic equilibrium \citep{Sarazin1988}; in turn, these masses can be used to infer constraints on cosmological parameters \citep[e.g.][]{Allen2011}. The temperature structure yields information on the detailed physics of shock-heated gas in merging events, the nature of cold fronts, and the role of turbulence and gas sloshing \citep[see e.g.][for a review]{mar07}. In turn, such analyses provide insights into the assembly physics of galaxy clusters, which is necessary to interpret the scaling relations between clusters masses and their primary observables \citep{Khedekar2013}.

However, the X-ray gas temperature measurement is potentially affected by two systematic effects. First, the X-ray emission is proportional to the square of the ICM electron density, such that spectroscopic temperatures are driven by the colder, denser regions along the line of sight and are thus sensitive to gas clumping. In fact, a weighted mean temperature is measured, in which the weight is a non-linear combination of the temperature and density structure \citep[see e.g.][]{maz04,vik06b}. Numerical simulations support this view \citep[e.g.][]{Nagai2007,ras14}, but estimates of the magnitude of any bias due to this effect vary widely depending on the numerical scheme (e.g. smoothed particle hydrodynamics, adaptive mesh refinement) and the details of sub-grid physics (cooling, feedback, etc). Secondly, the spectroscopic temperatures depend directly on the energy calibration of X-ray observatories. For instance, X-ray temperatures obtained with \textit{Chandra} are generally higher than those measured by XMM-\textit{Newton} by up to a factor of 15\% at 10~keV \citep[e.g.][]{Mahdavi2013}.

The thermal Sunyaev-Zel'dovich \citep[tSZ;][]{Sunyaev1972} effect is related to the mean gas-mass-weighted temperature along the line of sight and the electron density, via the ideal gas law. The tSZ effect can thus be used to obtain an alternative estimate of the gas temperature provided that a measure of the density is available. A combination of the tSZ and X-ray observations can then in principle be used to decouple temperature and density in each individual measurement. Such a method has previously been used to extract 1D gas temperature profiles, complementing X-ray spectroscopic measurements \citep[e.g.][]{Pointecouteau2002,Kitayama2004,Nord2009,Basu2010,Eckert2013,Ruppin2016}.

Here, we explore the application of the method to 2D data. We use deep, resolved ($< 20\arcsec$) tSZ observations, combined with X-ray imaging, to measure the spatial distribution of the gas temperature towards the merging cluster \mbox{MACS~J0717.5+3745} at $z=0.55$. We chose \mbox{MACS~J0717.5+3745} as a test case cluster because it is one of the very few objects for which tSZ data of sufficient depth and resolution are currently available \citep{Adam2016b}. The complex morphology of the cluster is the primary limiting factor to our analysis; however the system allows us to explore a wide range of gas temperatures, which are not necessarily accessible with more simple objects. We compare our new temperature map, based on X-ray and tSZ imaging, to that obtained from application of standard X-ray spectroscopic techniques using XMM-\textit{Newton} and \textit{Chandra} data. We describe and discuss in detail the various factors affecting the ratio between the two temperature estimates.
We assume a flat $\Lambda$CDM cosmology according to the latest {\it Planck} results \citep{Planck2015XIII} with $H_0 = 67.8$ km s$^{-1}$ Mpc$^{-1}$, $\Omega_M = 0.308$, and $\Omega_{\Lambda} = 0.692$. At the cluster redshift, 1 arcsec corresponds to 6.6 kpc.

\section{Data}\label{sec:data}
The New IRAM KIDs Array \citep[NIKA; see][]{Monfardini2011,Calvo2013,Adam2014,Catalano2014} has observed \mbox{MACS~J0717.5+3745} at 150 and 260 GHz for a total of 47.2 ks. The main steps of the data reduction are described in \cite{Adam2015,Adam2016a,Adam2016b,Ruppin2016}. In this paper, we use the NIKA 150 GHz tSZ map at 22 arcsec effective angular resolution full width half maximum (FWHM), deconvolved from the transfer function except for the beam smoothing. The overall calibration uncertainty is estimated to be 7\%, including the brightness temperature model of our primary calibrator, the NIKA bandpass uncertainties, the opacity correction, and the stability of the instrument \citep{Catalano2014}. The absolute zero level for the brightness on the map remains unconstrained by NIKA. \mbox{MACS~J0717.5+3745} is contaminated by a significant amount of kinetic SZ \citep[kSZ;][]{Sunyaev1980} signal and we used the best-fit model F2 from \cite{Adam2016b} to remove its contribution. This model has large uncertainties but it still allows us to test the impact of the kSZ effect on our results.

\mbox{MACS~J0717.5+3745} was observed several times by the XMM-\textit{Newton} and \textit{Chandra} X-ray observatories (obs-IDs 0672420101, 0672420201, 067242030, and 1655, 4200, 16235, 16305, respectively). The data processing follows the description given in \cite{Adam2016b}. The clean exposure time is 153 ks for \textit{Chandra} and 160 and 116 ks for XMM-\textit{Newton} MOS1,2 and PN cameras, respectively.

\section{Temperature reconstruction}\label{sec:method}
The method employed to recover the temperature of the gas from NIKA tSZ and XMM-\textit{Newton} X-ray imaging, $\TSZ$, is described below. The X-ray spectroscopic temperature mapping method is discussed in Section~\ref{sec:Xray_spectroscopic_temperature_map} and Appendix~\ref{append:Txerror}.

\subsection{Primary observables}
The tSZ signal, measured at frequency $\nu$, can be expressed as
\begin{equation}
        \frac{\Delta I_{\nu}}{I_0} = f(\nu, T_{\rm e}) \frac{\sigma_{\rm T}}{m_{\rm e} c^2} \int P_{\rm e} dl \equiv k_{\rm B} \TMW f(\nu, T_{\rm e}) \frac{\sigma_{\rm T}}{m_{\rm e} c^2} \int n_{\rm e} dl,
\label{eq:dIsz}
\end{equation}
where $f(\nu, T_{\rm e})$ is the tSZ spectrum, which depends slightly on temperature $T_{\rm e}$ in the case of very hot gas. The signal is proportional to the line of sight integrated electron pressure, $P_{\rm e}$. It is related to the mean gas-mass-weighted temperature along the line of sight, 
\begin{equation}
       \TMW  \equiv \frac{\int T_{\rm e} n_{\rm e} dl}{\int n_{\rm e} dl},
        \label{eq:Tgmw_def}
\end{equation}
and the electron density, $n_{\rm e}$, via the ideal gas law. The X-ray surface brightness is driven by the electron density
\begin{equation}
        S_{\rm X}= \frac{1}{4 \pi \left(1+z\right)^4} \int n_{\rm e}^2 \Lambda(T_{\rm e}, Z) \ dl,
        \label{eq:sx}
\end{equation}
where $z$ is the cluster redshift and $\Lambda(T_{\rm e}, Z)$ is the emissivity in the relevant energy band, taking into account the interstellar absorption and instrument spectral response. The parameter $\Lambda(T_{\rm e}, Z)$ depends only weakly on the temperature and metallicity of the gas $Z$, so that instrumental systematics have a negligible impact on the results presented in this paper.

\begin{figure*}[h]
\centering
\includegraphics[height=7.6cm]{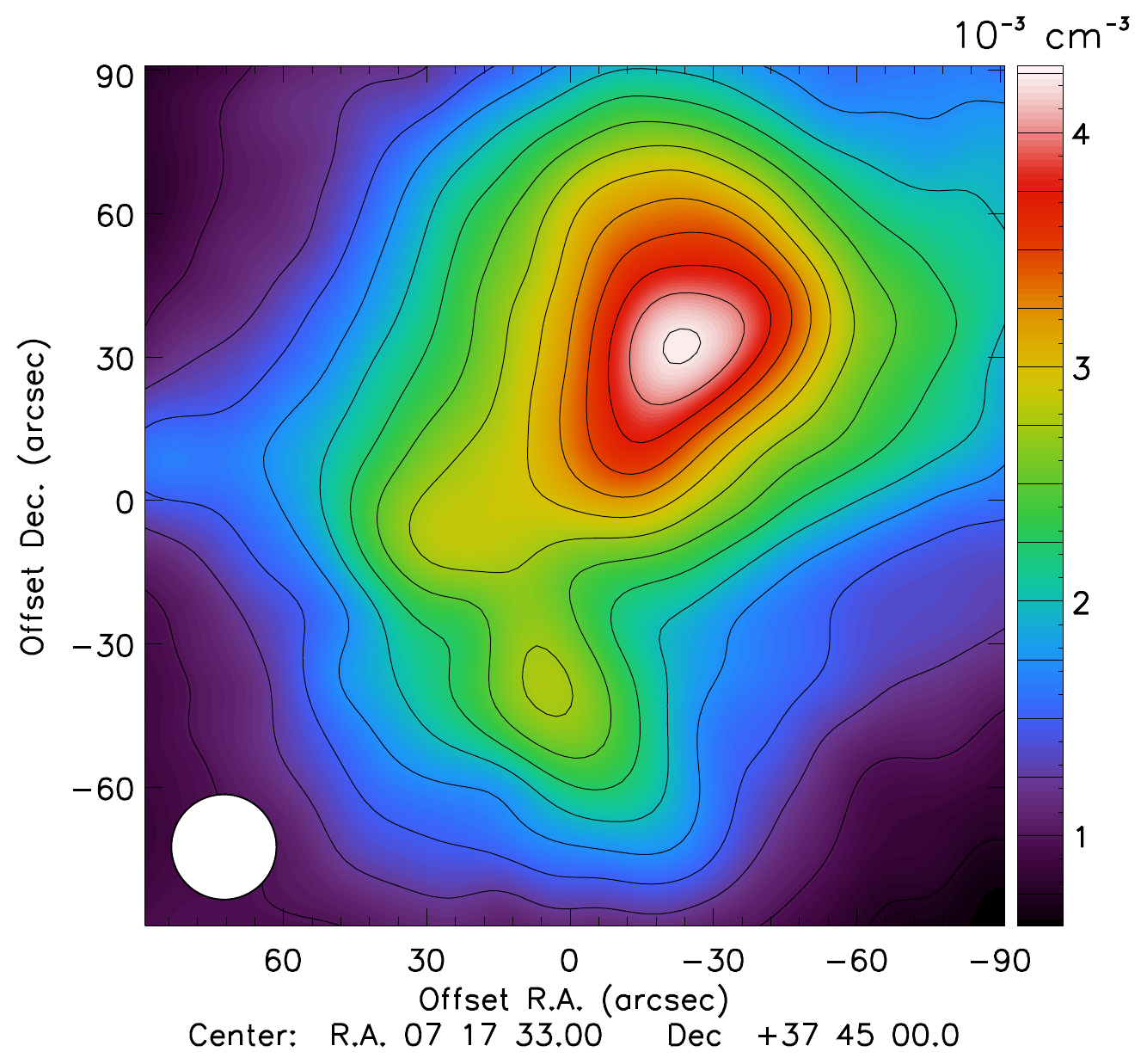}
\includegraphics[height=7.6cm]{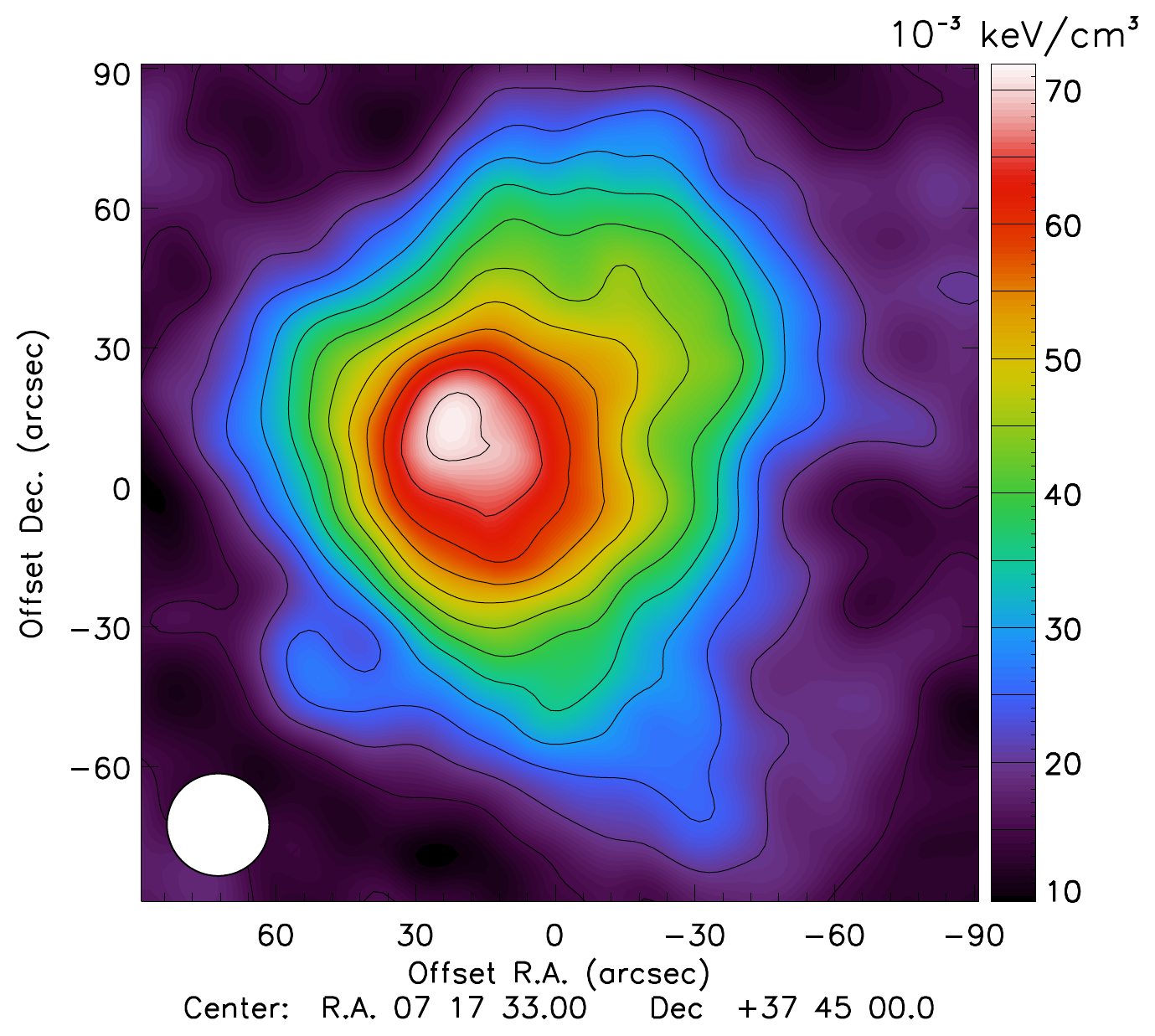}
\caption{\footnotesize{{\bf Left}: Effective line of sight electron density, $\overline{n}_{\rm e}$, derived from XMM-\textit{Newton}. {\bf Right}: Effective line of sight pressure, $\overline{P}_{\rm e}$, derived from NIKA is shown. These maps correspond to model M1, and were smoothed with a Gaussian kernel to an effective resolution of 22 arcsec FWHM. The pressure map is cleaned from our best-fit kSZ model and corrected for the zero level.}}\label{fig:Input_maps}
\end{figure*}

\subsection{X-ray electron density mapping}
We used the XMM-\textit{Newton} X-ray surface brightness (equation \ref{eq:sx}) to produce a map of the square of the electron density integrated along the line of sight, $\int n_{\rm e}^2 dl$. To combine it with tSZ observations, we had to convert $\int n_{\rm e}^2 dl$ to $\int n_{\rm e} dl$ via an effective electron depth, expressed as
\begin{equation}
        \ell_{\rm eff} = \frac{\left(\int n_{\rm e} dl\right)^2}{\int n_{\rm e}^2 dl}.
\label{eq:l_eff}
\end{equation}
From equation~\ref{eq:sx}, the average density along the line of sight is then given by
\begin{equation}
        \overline{n}_{\rm e} = \frac{1}{\ell_{\rm eff}} \int n_{\rm e} dl = \frac{1}{\sqrt{l_{\rm eff}}}\sqrt{\frac{4 \pi \left(1+z\right)^4 S_{\rm X}}{\Lambda\left(T_{\rm e}, Z\right)}},
\end{equation}
defining an effective density.

\subsection{Thermal Sunyaev-Zel'dovich pressure mapping}
Similarly, we can express the effective pressure along the line of sight directly from equation \ref{eq:dIsz}, as
\begin{equation}
        \overline{P}_{\rm e} = \frac{1}{\ell_{\rm eff}} \int P_{\rm e} dl = \frac{m_{\rm e} c^2}{\sigma_{\rm T}} \frac{y_{\rm tSZ}}{\ell_{\rm eff}}.
\label{eq:pseudo_p}
\end{equation}
We obtained this quantity in straightforward way from the NIKA map accounting for relativistic corrections as detailed in \cite{Adam2016b}. As the temperature can be very high, the relativistic corrections are non-negligible \citep{Pointecouteau1998,Itoh2003}, but the exact choice of the temperature map used to apply relativistic corrections has a negligible impact on our results (i.e. the spectroscopic temperature maps from XMM-\textit{Newton}, \textit{Chandra}, or $\TSZ$).

\subsection{Gas-mass-weighted temperature mapping}\label{sec:Gas_mass_weighted_temperature_mapping}
We obtained the tSZ+X-ray imaging temperature map, $\TSZ$, by combining the effective density and pressure 
\begin{equation}
        k_{\rm B} \TSZ 
        = \frac{\overline{P}_{\rm e}}{\overline{n}_{\rm e}} = \frac{1}{\sqrt{\ell_{\rm eff}}} \frac{m_{\rm e} c^2}{\sigma_{\rm T}} \sqrt{\frac{\Lambda\left(T_{\rm e}, Z\right)}{4 \pi \left(1+z\right)^4 S_{\rm X}}} y_{\rm tSZ}.
\label{eq:pseudo_tsz}
\end{equation}
The temperature map $\TSZ$ is an estimate of the gas-mass-weighted temperature, $\TMW$ (equation~\ref{eq:Tgmw_def}). We propagated the noise arising from the tSZ map and the X-ray surface brightness with Monte Carlo realisations; the overall noise on $\TSZ$ is dominated by that of the tSZ map. The sources of systematic errors are incorrect modelling of $\ell_{\rm eff}$ along with tSZ calibration uncertainties and contamination from the kSZ effect. The absolute calibration error of the X-ray flux is expected to be negligible.

\begin{figure*}[h]
\centering
\includegraphics[trim=0cm 0cm 1.4cm 0cm, clip=true, totalheight=7.6cm]{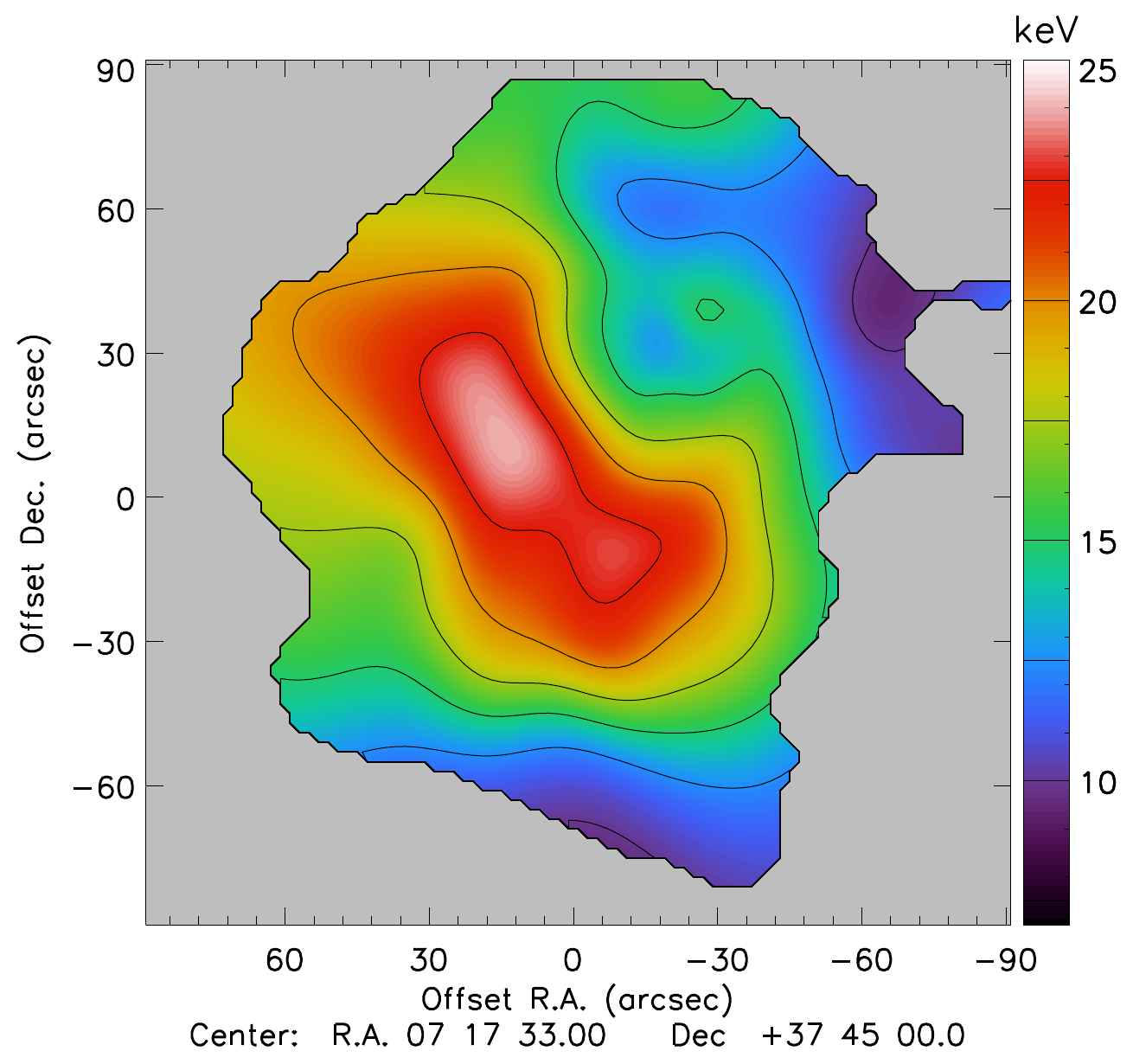}
\includegraphics[trim=0cm 0cm 0cm 0cm, clip=true, totalheight=7.6cm]{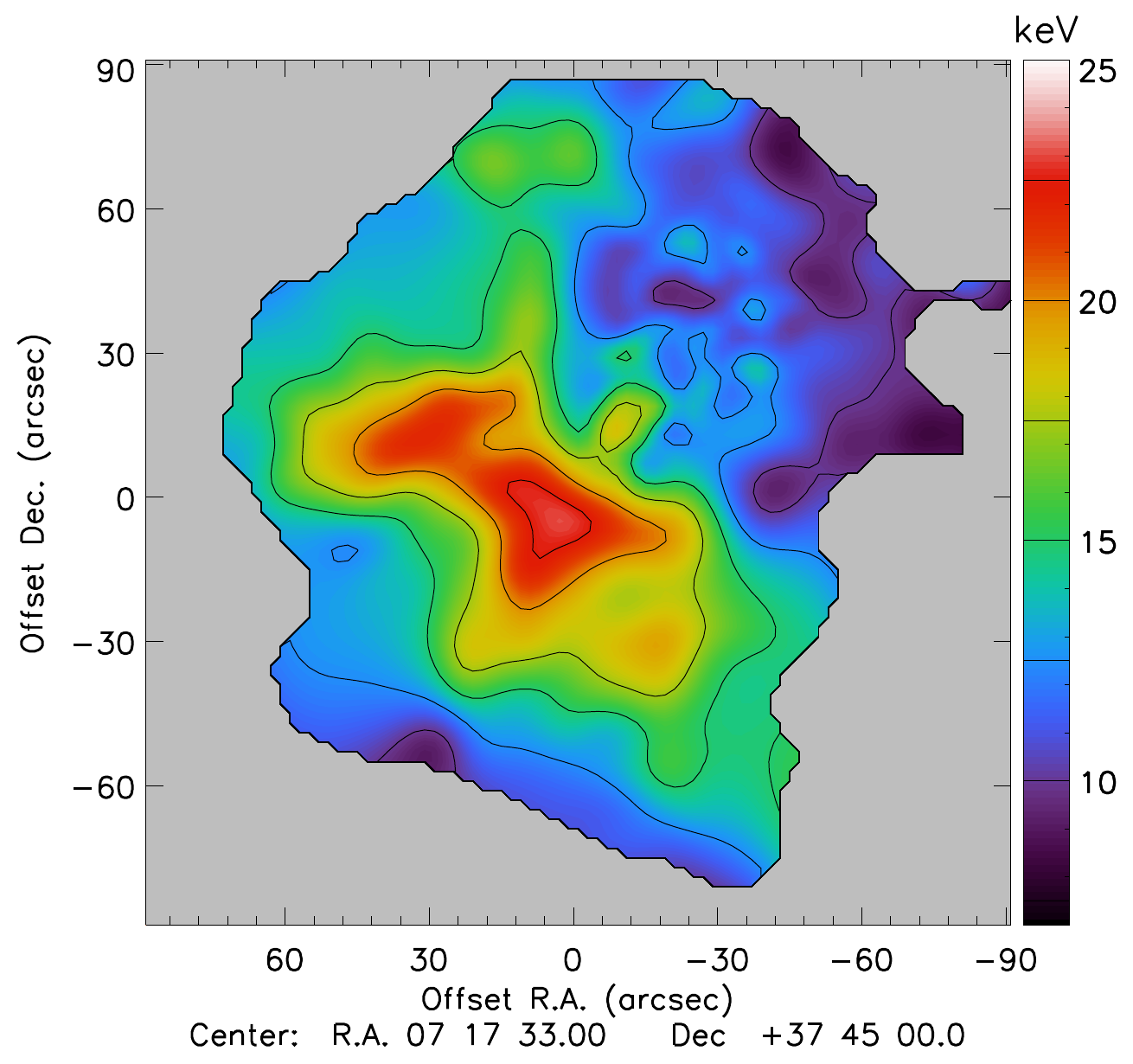}
\includegraphics[trim=0cm 0cm 1.4cm 0cm, clip=true, totalheight=7.6cm]{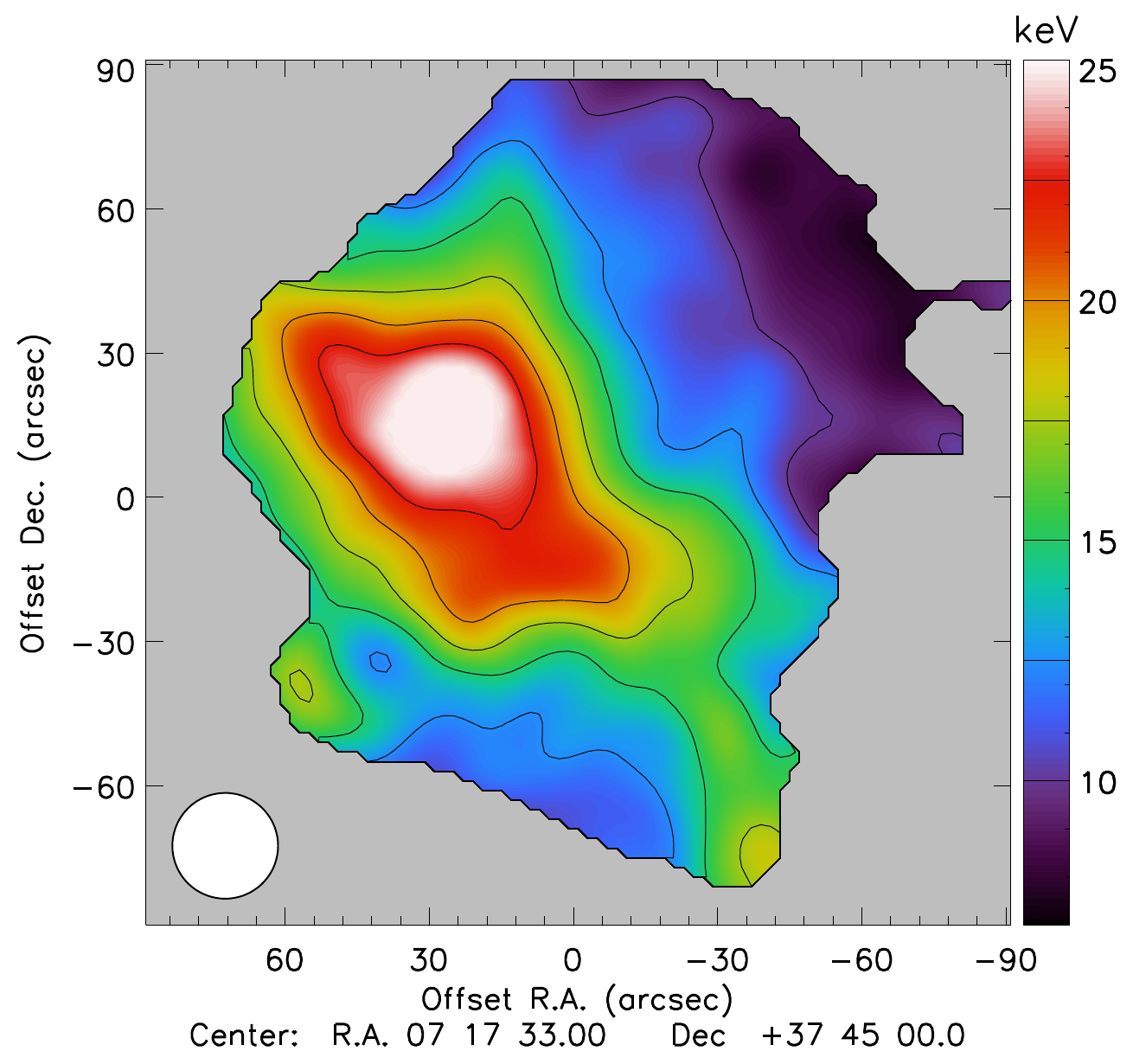}
\includegraphics[trim=0cm 0cm 0cm 0cm, clip=true, totalheight=7.6cm]{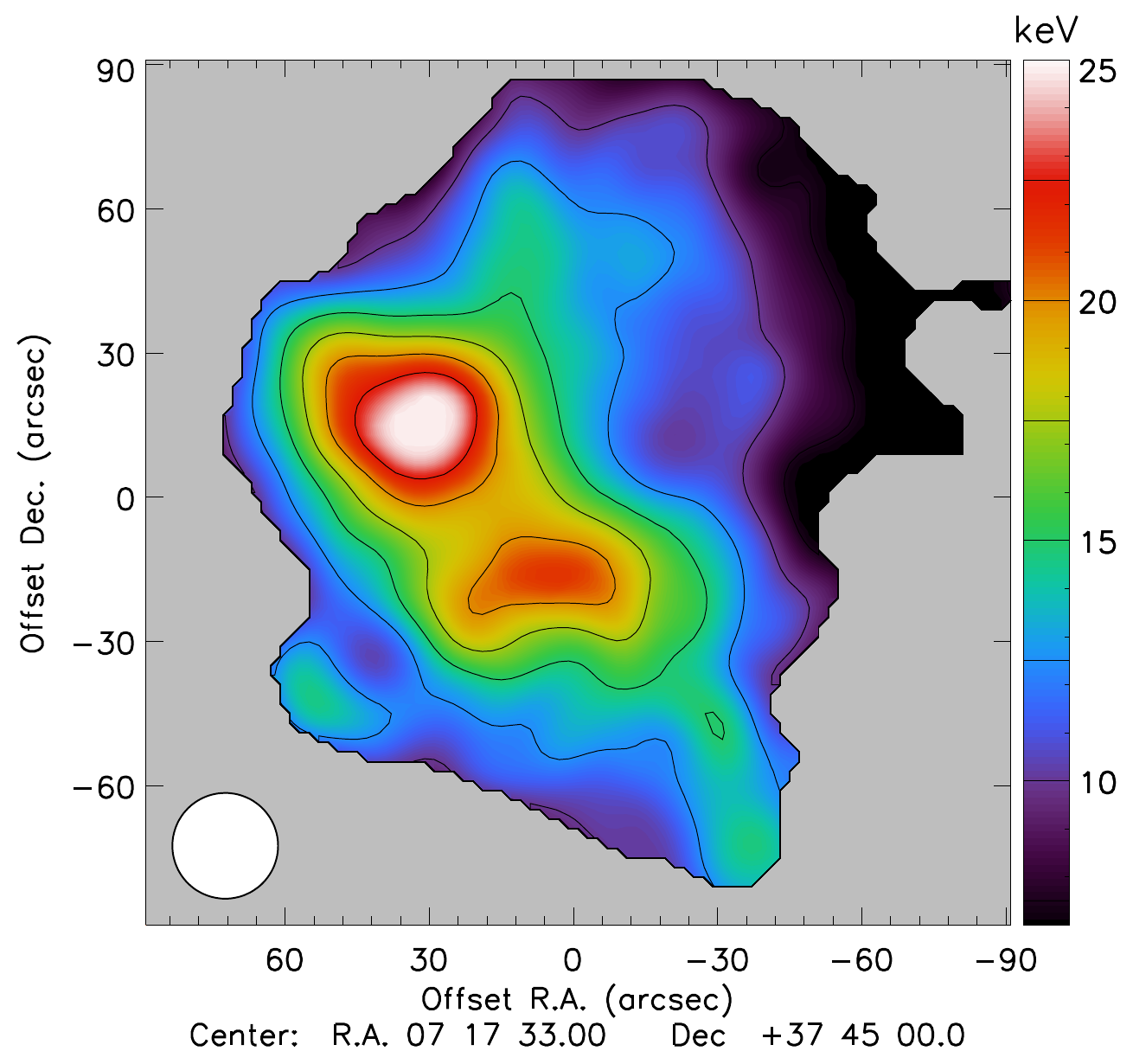}
\caption{\footnotesize{Temperature maps. {\bf Top}: Spectroscopic temperature derived from \textit{Chandra} ($\TXC$, left panel) and from XMM-\textit{Newton} ($\TXX$, right panel) are shown. {\bf Bottom}: NIKA and XMM \textit{Newton} imaging derived temperature, $\TSZ$, for model M1 (left panel) and model M3a (right panel) are shown. These maps are corrected for the zero level.}}
\label{fig:T_maps}
\end{figure*}

\subsection{Effective electron depth}
The effective electron depth is a key quantity for the method, as the derived gas-mass-weighted temperature scales with $\sqrt{\ell_{\rm eff}}$. It can be re-expressed as
\begin{equation}
\ell_{\rm eff} = \frac{R_{500}}{Q_{n_{\rm e}}^2}~~{\rm with}~~Q_{n_{\rm e}} = \frac{ \sqrt{\langle \xe^2\rangle}}{ \langle\xe\rangle},
\label{eq:Q_ne}
\end{equation}
where the brackets denote averaging along the line of sight, carried out in scaled coordinates. The electron depth at each projected position depends, via the shape factor $Q_{n_{\rm e}}$, on the geometry of the gas density distribution at all scales, from the large-scale radial dependence to small-scale fluctuations. In particular, $Q_{n_{\rm e}}$ increases with increasing gas concentration and clumpiness.

In the following, we used several approaches to estimate $\ell_{\rm eff}$ and its uncertainty:
\begin{enumerate}
\item {\bf Model M1:} Following \cite{Sayers2013}, we assumed that $\ell_{\rm eff}$ is constant at $\ell_{\rm eff}=1400$ kpc, as estimated by \citet{Mroczkowski2012}, across the cluster extension. 
\item {\bf Model M2:} We derived an electron density profile from deconvolution and deprojection of the XMM-\textit{Newton} radial $S_{\rm X}$ profile centred on the X-ray peak \citep{Croston2006}, thus obtaining an azimuthally symmetric $\ell_{\rm eff}$ map. 
\item {\bf Model M3:} We used the best-fitting NIKA tSZ and XMM-\textit{Newton} density model of \cite{Adam2016b}, which accounts for the four main subclusters in \mbox{MACS~J0717.5+3745}, to compute a map of $\ell_{\rm eff}$. The model does not constrain the line of sight distance between the subclusters because the tSZ signal depends linearly on the density. Therefore, we considered two extreme cases: {\bf M3a)} where the subclusters are sufficiently far away from each other such that $\int n_{\rm e}^2 dl \simeq \sum_j \int n_{{\rm e},j}^2 dl$, where $j$ refers to each subcluster; {\bf M3b)} where all the subclusters are located in the same plane, perpendicular to the line of sight. The physical distances between the subclusters are thus minimal, maximizing the $\int n_{\rm e}^2 dl$ integral.
\end{enumerate}
While the internal structure of \mbox{MACS~J0717.5+3745} is increasingly refined from model M1 to M3, we found good consistency between all three models. Model M2 presents a minimum of 1200 kpc towards the X-ray centre and increases quasi-linearly towards higher radii, reaching about 2000 kpc at 1 arcmin, in line with expectations from model M1. Model M3a is minimal in the central region in the direction of the subclusters ($\sim 1200$ kpc) and also increases with radius. Model M3b provides a lower limit for $\ell_{\rm eff}$, increasing from $\sim 800$ kpc near the centre to $\sim 1200$ kpc at 1 arcmin.

While these models allowed us to test the impact of the gas geometry on large scales, they do not specifically account for clumping on small scales. Despite the weak dependence of the gas-mass-weighted temperature on the electron depth ($\propto \sqrt{\ell_{\rm eff}}$), clumping might affect our results. We discuss this further in Section \ref{sec:results}.

\begin{figure*}[h]
\centering
\includegraphics[width=0.32\textwidth]{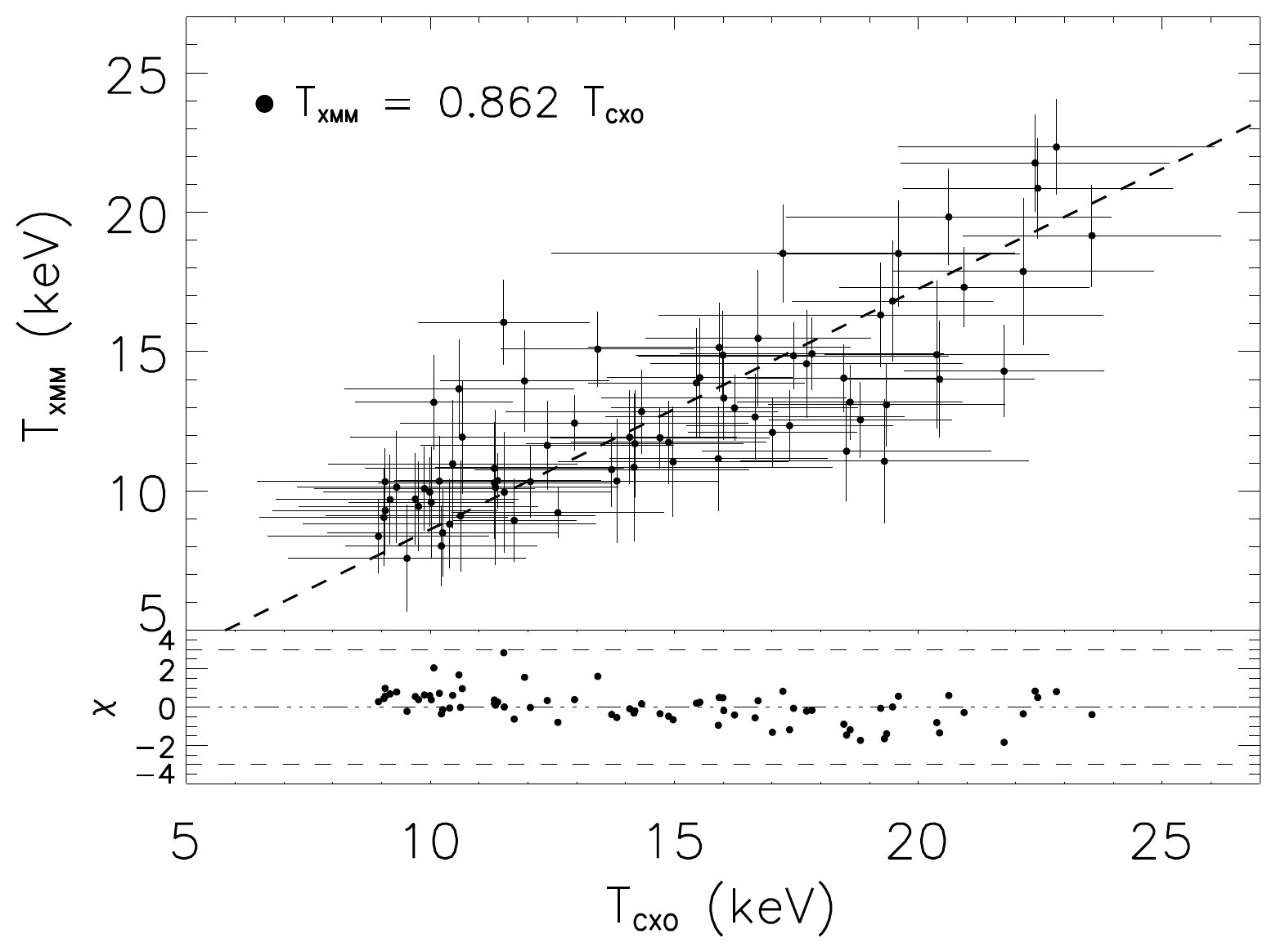}
\includegraphics[width=0.32\textwidth]{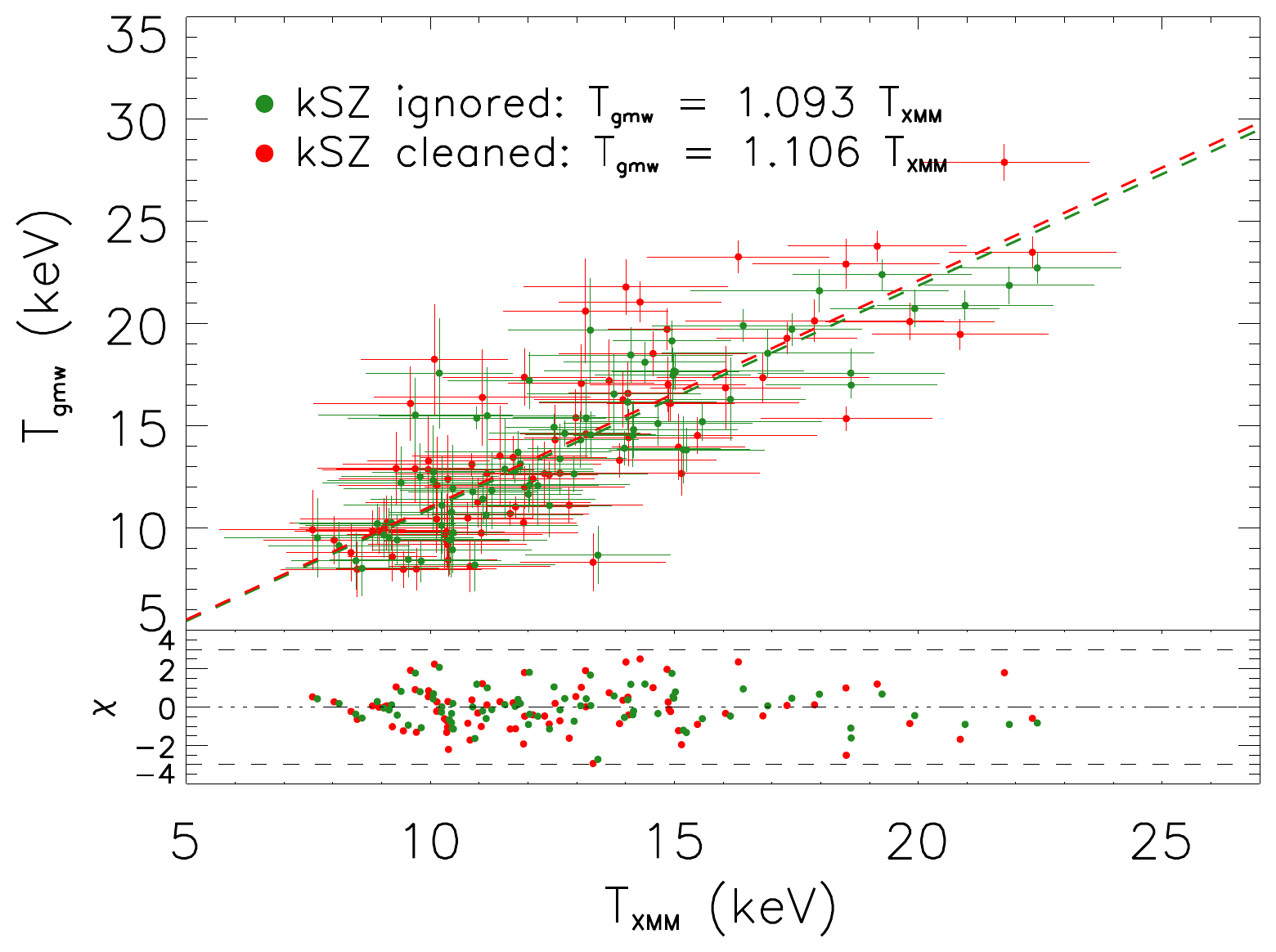}
\includegraphics[width=0.32\textwidth]{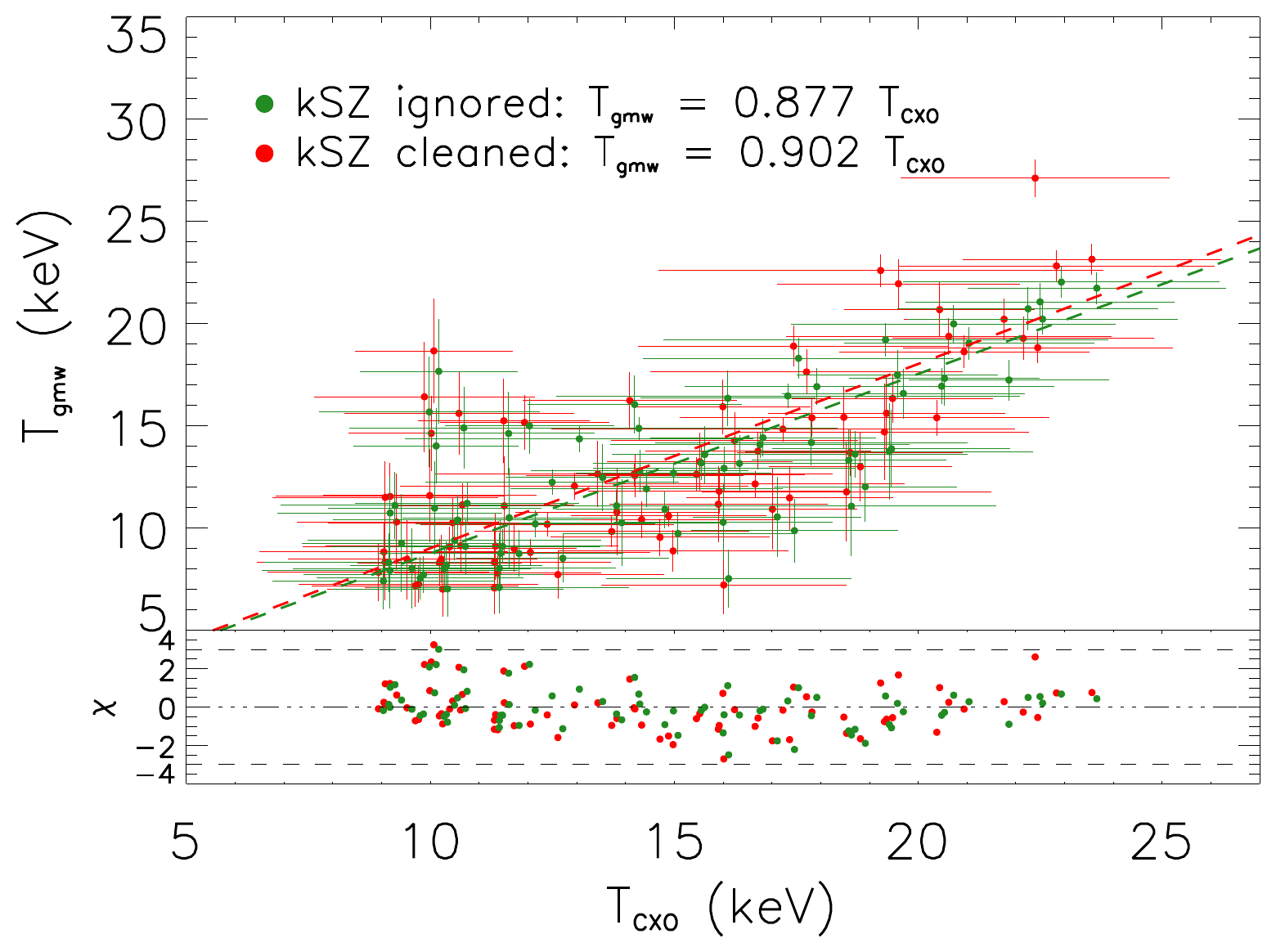}
\caption{\footnotesize{Correlation between the temperature maps of Figure \ref{fig:T_maps} and residual. {\bf Left:} XMM-\textit{Newton} vs. \textit{Chandra} spectroscopic temperatures is shown. {\bf Middle:} tSZ+X-ray imaging (model M1) vs. XMM-\textit{Newton} spectroscopy is shown. {\bf Right:} tSZ+X-ray imaging (model M1) vs. \textit{Chandra} spectroscopy is shown. The red and green dots correspond to the case with and without the kSZ correction, respectively.}}
\label{fig:T_SZ_T_X_correlation}
\end{figure*}

\subsection{X-ray spectroscopic temperature maps}\label{sec:Xray_spectroscopic_temperature_map}
The X-ray spectroscopic temperature maps from \textit{Chandra} (${T}_{\rm CXO}$) and XMM-\textit{Newton} (${T}_{\rm XMM}$) were produced using the wavelet filtering algorithm described in \cite{Bourdin2008}, as detailed in \cite{Adam2016b}. As the significance of wavelet coefficients partly depends on the photon count statistics, the effective resolution varies across the map, XMM-\textit{Newton} allowing a finer sampling than \textit{Chandra} owing to a higher effective area. We estimated the uncertainties per map pixel using a Monte Carlo approach, as discussed in Appendix~\ref{append:Txerror}.

Comparison of the temperature derived from tSZ+X-ray imaging to the X-ray spectroscopic temperature provides further information on the ICM structure and on calibration systematics. From equations~\ref{eq:pseudo_tsz} and \ref{eq:Q_ne}, at each projected position the ratio of the two temperatures can be expressed as
\begin{equation}
        \frac{\TSZ }{\TX} = \frac{\TMW}{T_{\rm spec}}\,\frac{Q_{n_{\rm e}, \ {\rm model}}} {Q_{n_{\rm e}, \ {\rm true}}} \, C_{\rm X}\,C_{\rm SZ},
\label{eq:T_ratio}
\end{equation}
where $\TMW$ is the true gas-mass-weighted temperature and $T_{\rm spec}$ is the spectroscopic temperature that would be obtained by fitting an isothermal model to the observed spectra for a perfectly calibrated instrument. The value $C_{\rm X} = T_{\rm spec}/\TX$ is the ratio between the latter and the measured X-ray temperature, which accounts for the X-ray calibration uncertainty. The value $C_{\rm SZ}$ accounts for the tSZ calibration.

The measured ratio $\TSZ/\TX$ is an estimate of the ratio between the gas-mass-weighted temperature and the spectroscopic temperature, $Q{_{\rm T}}= \TMW/T_{\rm spec}$. The spectroscopic temperature $T_{\rm spec}$ is expected to be biased low as compared to the gas-mass-weighted temperature and depends on the instrument used to make the measurement. The ratio $Q_{\rm T}$ is a shape parameter, which depends on both the density and temperature structure along the line of sight. In addition to calibration issues, the measured ratio $\TSZ/\TX$ may differ from $Q{_{\rm T}}$ if the density shape factor, $Q_{n_{\rm e}}$, is incorrect. For a given cluster, the various terms on the right-hand side of equation~\ref{eq:T_ratio} are in principle degenerate. Part of the degeneracy, in particular of calibration versus physical factors, can be broken by taking into account the expected differences in spatial dependence.

\section{Results}\label{sec:results}
\subsection{Morphology}
The left and right panels of Figure \ref{fig:Input_maps} represent the effective density and pressure maps in the case of the simplest model M1, thus $\propto \sqrt{\int\xe^2 dl}$ and $\propto \int \pe dl$, respectively. The pressure map is corrected for the kSZ and the zero level (see Section \ref{sec:compT}). The cluster clearly exhibits a disturbed morphology. The morphology of the ICM pressure is similar to that of the density on large scales, but we observe strong differences at the substructure level, indicating spatial variations of the temperature. In particular, the pressure peak is offset $\sim 30$ arcsec south-east with respect to the density peak.

Figure \ref{fig:T_maps} shows the temperature maps ${T}_{\rm CXO}$, ${T}_{\rm XMM}$, and $\TSZ$ for models M1 and M3a. $\TSZ$ is corrected for the zero level and kSZ-corrected. All the maps identify a hot gas bar to the south-east. The position of the temperature peak is the same for $\TXC$ and $\TSZ$, while it is slightly shifted south-west for $\TXX$; however it also coincides with a region where kSZ contamination is large, leading to possible overestimation in $\TSZ$. All four maps indicate cooler temperatures in the the north-west sector. Varying the kSZ correction and the $\ell_{\rm eff}$ models slightly changes the local morphology of $\TSZ$ in the bar. Use of model M3a leads to the appearance of a secondary peak, while there is also a hint of a bimodal bar structure in the X-ray spectroscopic maps. However, the general agreement with the X-ray spectroscopic results, both in terms of absolute temperature and morphology, is good in all cases. 

\subsection{Temperature comparison}\label{sec:compT}
Figure \ref{fig:T_SZ_T_X_correlation} shows the correlation between the maps shown in Figure \ref{fig:T_maps}. Both tSZ+X-ray and X-ray spectroscopic temperature values were extracted in 20 arcsec pixels (see Appendix~\ref{append:correlation} for details). We masked pixels, where the tSZ signal-to-noise ratio ${\rm S/N} < 2$, to avoid possible bouncing effects on the edge of the map due to the NIKA data processing.

\begin{figure*}[h]
\centering
\includegraphics[width=0.49\textwidth]{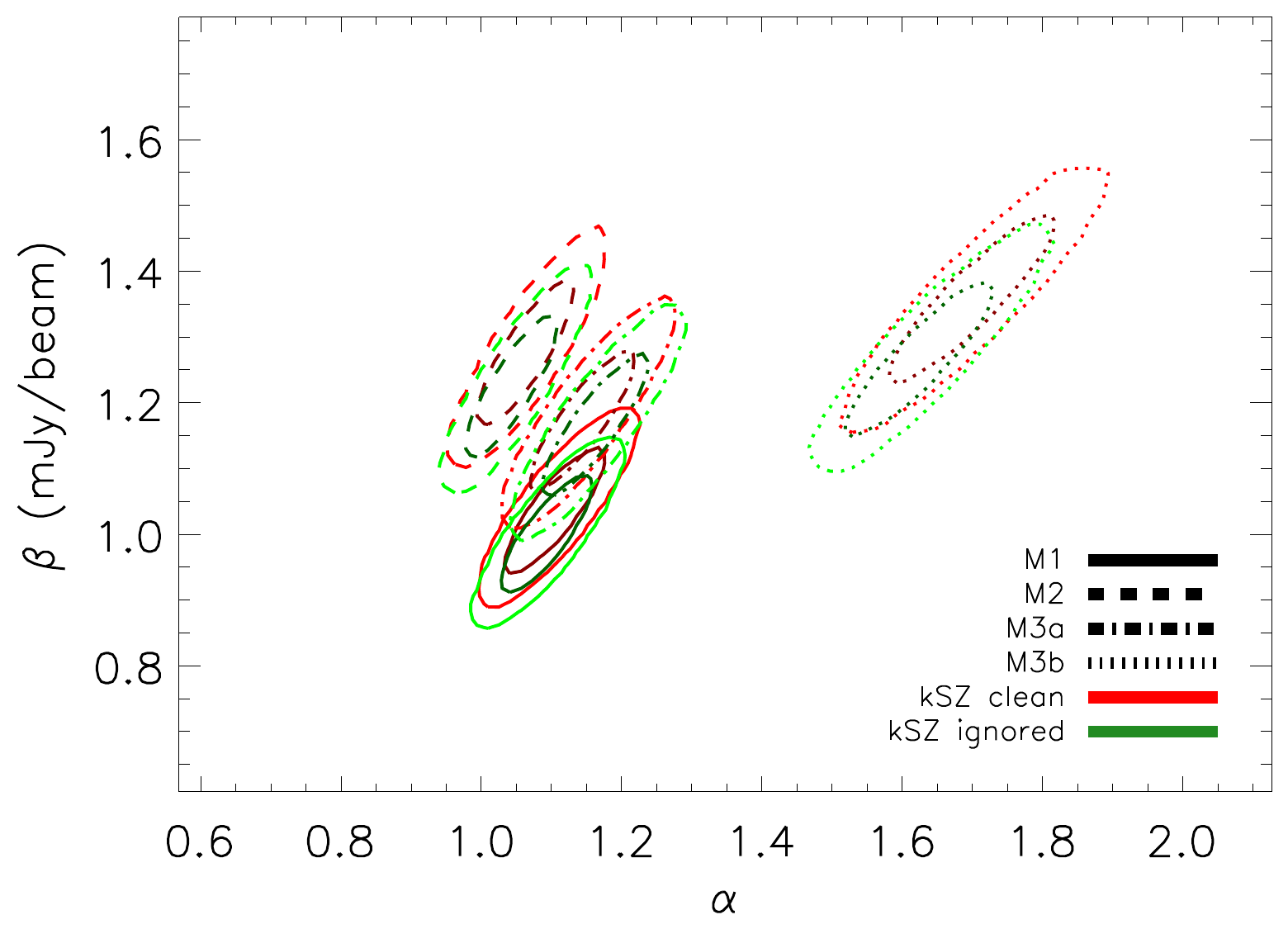}
\includegraphics[width=0.49\textwidth]{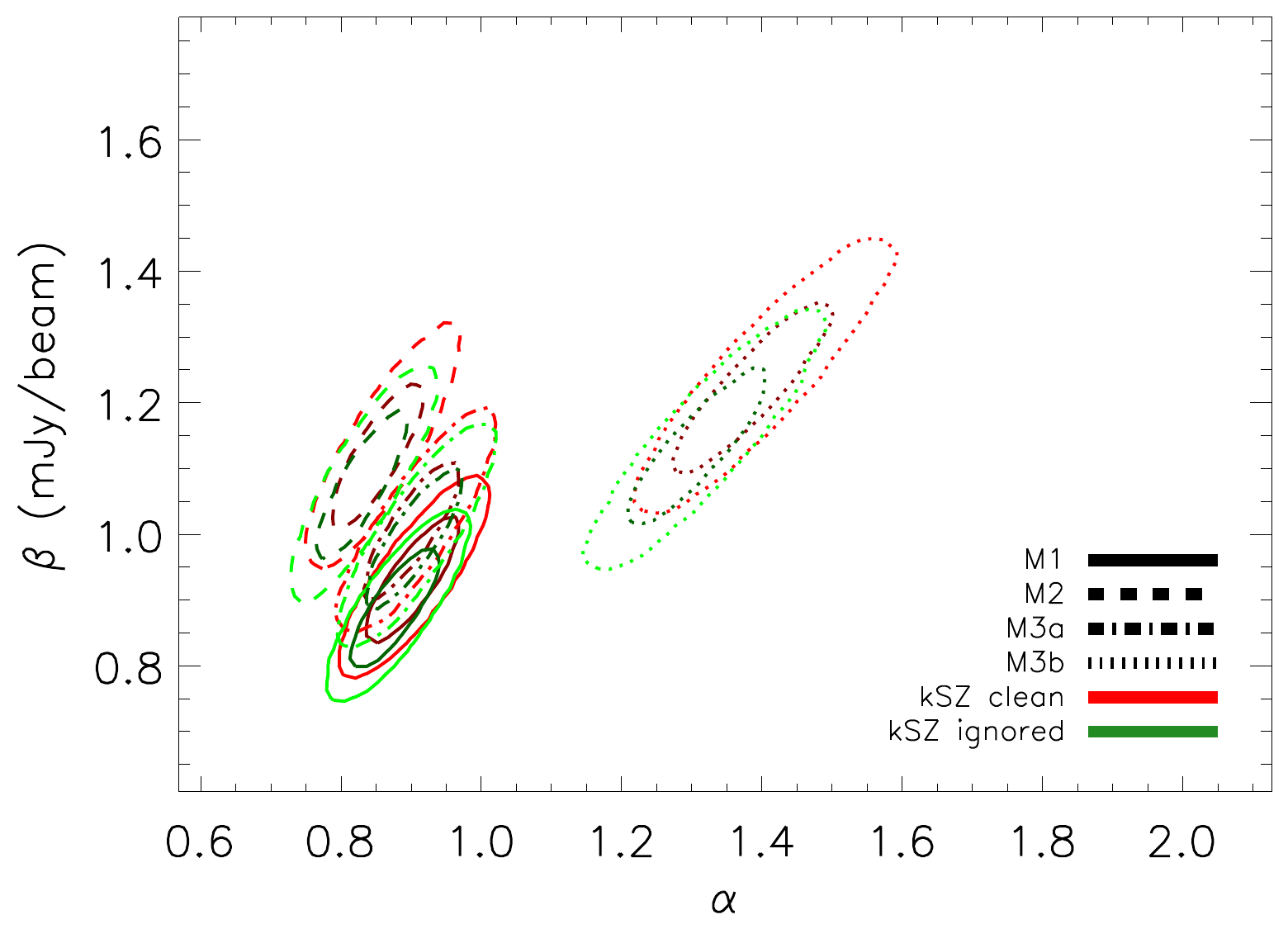}
\caption{\footnotesize Posterior likelihood (68 and 95\% C.L.) in the plane $\alpha$ -- $\beta$ (expressed in term of the zero level brightness of the NIKA map). {\bf Left:} tSZ+X-ray imaging vs. XMM-\textit{Newton} spectroscopy is shown. {\bf Right:} tSZ+X-ray imaging vs. \textit{Chandra} spectroscopy is shown. The red and green dots correspond to the case with and without the kSZ correction, respectively, and the different models are provided with different dashed line as shown in the legend.}
\label{fig:T_SZ_T_X_posterior}
\end{figure*}

Since the zero level of the tSZ map is unconstrained, we express the effective pressure map as $\overline{P}_{\rm e} = \overline{P}_{\rm true} + \overline{P}_0$, where $\overline{P}_0$ is an unknown offset. Following equation \ref{eq:pseudo_tsz}, the gas-mass-weighted temperature can then be expressed with respect to the spectroscopic temperature as
\begin{equation}
k_{\rm B} \TSZ = \alpha_{\rm SZX} \times k_{\rm B} \TXXC+ {\beta}/{\overline{n}_{\rm e}},
\label{eq:TSZ_TX_regression}
\end{equation}
where $\beta$ gives a measurement of $\overline{P}_0$. For X-ray spectroscopic temperatures, we simply write $\TXX = \alpha_{\rm XMM-CXO} \times \TXC$. We perform a linear regression between the pairs of temperature maps accounting for error bars on both axis, as detailed in Appendix~\ref{append:correlation}. Table~\ref{tab:regression_coeff} gives the $\alpha $ and $\beta$ coefficients and the intrinsic scatter, obtained for the different $\ell_{\rm eff}$ models tested, and their dependence on the kSZ correction. Figure \ref{fig:T_SZ_T_X_posterior} provides the posterior likelihood in the $\alpha_{\rm SZX}$ -- $\beta$ plane for all the regressions performed between $\TSZ$ and ${T}_{\rm XMM/CXO}$.

\begin{table*}[]
\caption{\footnotesize{Regression and intrinsic scatter coefficients between the temperature maps. The central value is the median of the posterior likelihood and the errors are obtained by integrating the posterior likelihood within 90\% C.L. The posterior likelihood distribution is highly non-Gaussian in the case of the scatter and error bars should be interpreted with caution.} $^{\star}$Model M3b gives a lower limit for $\ell_{\rm eff}$, and thus should be taken only as an upper limit for $\alpha$.}
\begin{center}
\resizebox{\textwidth}{!} {
\begin{tabular}{c|ccc|c}
\hline
\hline
 & \multicolumn{4}{c}{$\ell_{\rm eff}$ model} \\
Slope / offset (mJy/Beam) / scatter (keV) & M1 & M2 & M3a & M3b$^{\star}$ \\
\hline
 & \multicolumn{4}{c}{kSZ-corrected} \\
\hline
$\left( \alpha, \beta, \sigma_{\rm int} \right)_{\rm SZX-XMM}$ & $\left(1.11_{-0.07}^{+0.08} , 1.04_{-0.10}^{+0.10} , 1.43_{-0.62}^{+0.38}\right)$ & $\left(1.06_{-0.07}^{+0.07} , 1.28_{-0.12}^{+0.12} , 1.29_{-0.60}^{+0.35}\right)$ & $\left(1.15_{-0.08}^{+0.08} , 1.17_{-0.11}^{+0.12} , 1.59_{-0.55}^{+0.37}\right)$ & $\left(1.70_{-0.12}^{+0.13} , 1.36_{-0.14}^{+0.14} , 2.44_{-0.71}^{+0.50}\right)$ \\
$\left( \alpha, \beta, \sigma_{\rm int} \right)_{\rm SZX-CXO}$ & $\left(0.90_{-0.07}^{+0.07} , 0.93_{-0.10}^{+0.11} , 1.59_{-0.78}^{+0.47}\right)$ & $\left(0.85_{-0.06}^{+0.07} , 1.12_{-0.12}^{+0.13} , 1.51_{-0.67}^{+0.41}\right)$ & $\left(0.90_{-0.07}^{+0.08} , 1.01_{-0.11}^{+0.12} , 2.51_{-0.40}^{+0.36}\right)$ & $\left(1.39_{-0.11}^{+0.14} , 1.23_{-0.13}^{+0.16} , 2.50_{-1.00}^{+0.64}\right)$ \\
\hline
 & \multicolumn{4}{c}{kSZ-uncorrected} \\
\hline
$\left( \alpha, \beta, \sigma_{\rm int} \right)_{\rm SZX-XMM}$ & $\left(1.09_{-0.07}^{+0.07} , 1.00_{-0.09}^{+0.10} , 0.00_{-0.00}^{+0.56}\right)$ & $\left(1.04_{-0.07}^{+0.07} , 1.23_{-0.11}^{+0.12} , 0.00_{-0.00}^{+0.61}\right)$ & $\left(1.16_{-0.08}^{+0.08} , 1.17_{-0.11}^{+0.12} , 1.51_{-0.61}^{+0.37}\right)$ & $\left(1.63_{-0.11}^{+0.12} , 1.27_{-0.12}^{+0.13} , 0.00_{-0.00}^{+0.69}\right)$ \\
$\left( \alpha, \beta, \sigma_{\rm int} \right)_{\rm SZX-CXO}$ & $\left(0.88_{-0.06}^{+0.07} , 0.89_{-0.09}^{+0.10} , 0.73_{-0.73}^{+0.71}\right)$ & $\left(0.83_{-0.06}^{+0.07} , 1.08_{-0.12}^{+0.12} , 0.82_{-0.82}^{+0.60}\right)$ & $\left(0.90_{-0.07}^{+0.08} , 1.00_{-0.11}^{+0.12} , 2.52_{-0.40}^{+0.35}\right)$ & $\left(1.31_{-0.10}^{+0.12} , 1.13_{-0.13}^{+0.14} , 0.60_{-0.60}^{+1.25}\right)$ \\
\hline
$\left( \alpha, \sigma_{\rm int} \right)_{\rm XMM-CXO}$ & \multicolumn{4}{c}{$\left(0.86_{-0.03}^{+0.03} , 0.00_{-0.00}^{+0.00}\right)$} \\
\hline
\end{tabular}
}
\end{center}
\label{tab:regression_coeff}
\end{table*}

The ratio of the temperature obtained from tSZ+X-ray imaging versus the temperature obtained from X-ray spectroscopy is stable to within 10\%, depending on the choice of the $\ell_{\rm eff}$ model and kSZ correction used. Model M3b provides a lower limit on $\ell_{\rm eff}$, and therefore an upper limit on $\alpha_{\rm SZX}$. The scatter of about 2 keV between $\TXC$ and $\TXX$ is dominated by the statistical error. The scatter between $\TSZ$ and both X-ray temperatures are comparable, but slightly lower for $\TXX$. In most cases, the scatter is compatible with the noise as propagated into the $\TSZ$ and $\TX$ maps. The intrinsic scatter is only detected significantly, at the $\sim 2-3\sigma$ level, for the model M3a. This may be due to a number of factors, including the difference in angular resolution of the maps or an intrinsic scatter between gas-mass-weighted and spectroscopic temperatures.

Figure~\ref{fig:T_SZ_T_X_correlation} and Table~\ref{tab:regression_coeff} indicate that \textit{Chandra} temperatures are about 15\% higher than those of XMM-\textit{Newton}, as found by previous work \citep{Mahdavi2013,sch15}, while $\TSZ$ is on average larger than $\TXX$ and lower than $\TXC$ by about 10\%. The reasonable agreement between $\TSZ$ and $\TX$ suggests that there is no major flaw in the method and/or unidentified systematic effects in the analysis.

When dealing with multiphase plasma, X-ray spectroscopic temperatures are expected to underestimate the gas temperature by 10-20\% \citep{Mathiesen2001,maz04}. This is particularly important in the presence of strong temperature gradients, as would be expected in strong mergers such as \mbox{MACS~J0717.5+3745}. We observe such a difference when comparing $\TSZ$ with the lower $\TXX$ values, but not with $\TXC$. This must not be over-interpreted in terms of X-ray calibration. First, the difference is not very significant, taking into account the statistical errors on the ratio ($\sim 7\%$, Table~\ref{tab:regression_coeff}) and the absolute calibration of the tSZ map, which is expected to be accurate to within 7\%. Furthermore, the gas clumpiness is not taken into account in the model. This would under-estimate the $Q_{n_{\rm e}}$ factor and thus the measured $\TSZ$ values (equation~\ref{eq:T_ratio}). For instance, combining Planck tSZ \citep{Planck2015I} and XMM-\textit{Newton} observations, \citet{Tchernin2016} have found the $Q_{n_{\rm e}}$ clumpiness factor to be about 10\% in the cluster Abell 2142 within 1~Mpc from the centre, increasing to about 20\% at $R_{200}$. Numerical simulations suggest a factor of up to $\sim 40$\% at $R_{200}$, but with a rather large cluster-to-cluster scatter \citep[e.g.][]{Nagai2011,Zhuravleva2013,Vazza2013}. A clumpiness factor of $20\%$ would put $\TSZ$ in better agreement with $\TXC$ values. This illustrates the difficulty in disentangling various instrumental effects and intrinsic cluster properties, especially on a single cluster with a particularly complex morphology.

\section{Conclusions}\label{sec:conclusions}
Using deep tSZ observations together with X-ray imaging, we have extracted an ICM temperature map of the galaxy cluster \mbox{MACS~J0717.5+3745}. This map is weighted by gas mass and provides an alternative to purely X-ray spectroscopic-based methods. Because the test cluster is extremely hot, with the peak temperature reaching up to $\sim 25$ keV, this allows us to sample a large range of temperature, which would not be accessible with the large majority of clusters.

The morphological comparison of the gas-mass-weighted temperature map to XMM-\textit{Newton} and \textit{Chandra} X-ray spectroscopic maps indicates good agreement between the different methods. All three maps are consistent with \mbox{MACS~J0717.5+3745} having a low temperature in the north-west region and presenting a bar-like high temperature structure to the south-east, which is indicative of heating from adiabatic compression owing to the merger between two main subclusters \citep[see e.g.][]{Ma2009}.

We performed a first quantitative comparison between the various maps. The ratio of the temperature obtained from tSZ+X-ray imaging versus the temperature obtained from X-ray spectroscopy is stable to within 10\%, depending on the choice of the large scale density model and the kSZ correction used. We found that \textit{Chandra} temperatures are about $15\%$ higher than those of XMM-\textit{Newton}, as found by previous work, while $\TSZ$ is on average higher than $\TXX$ and lower than $\TXC$ by about 10\% in each case. Such ratios are typical and are consistent with expectations, taking into account cluster structures and measurement systematics. The gas-mass-weighted temperature map we derived is limited by the complexity of the test cluster and by assumptions on the effective electron depth of the ICM, kSZ contamination, and the calibration of the NIKA instrument. For a perfectly spherical cluster, the ratio $T_{\rm X} / \TSZ$ would give access to absolute calibration of the X-ray temperature. Since clusters are complex objects, the ratio we really measure is a complicated combination of the 3D temperature structure and intrinsic properties affecting the density, such as the amount of substructure, gas clumpiness, and triaxiality. A larger sample would allow us to disentangle instrumental calibration from effects linked to intrinsic cluster properties.

The noise in our $\TSZ$ map is significantly lower, especially at high temperatures, to that obtained from XMM-\textit{Newton} and \textit{Chandra}, but obtained with a factor of three smaller observing time. This illustrates the potential of resolved tSZ observations at intermediate to high redshifts, where X-ray spectroscopy becomes challenging, and which should be routinely provided by the upcoming generation of SZ instruments, MUSTANG2 \citep{Dicker2014} and NIKA2 \citep{Calvo2016,Comis2016}.

\begin{acknowledgements}
We are thankful to the anonymous referee for useful comments that helped improve the quality of the paper.
We would like to thank the IRAM staff for their support during the campaigns. 
We thank Marco De Petris for useful comments.
The NIKA dilution cryostat has been designed and built at the Institut N\'eel. In particular, we acknowledge the crucial contribution of the Cryogenics Group, and in particular Gregory Garde, Henri Rodenas, Jean Paul Leggeri, Philippe Camus. 
This work has been partially funded by the Foundation Nanoscience Grenoble, the LabEx FOCUS ANR-11-LABX-0013 and the ANR under the contracts "MKIDS", "NIKA" and ANR-15-CE31-0017. 
This work has benefited from the support of the European Research Council Advanced Grants ORISTARS and M2C under the European Union's Seventh Framework Programme (Grant Agreement nos. 291294 and 340519).
We acknowledge funding from the ENIGMASS French LabEx (B. C. and F. R.), the CNES post-doctoral fellowship programme (R. A.), the CNES doctoral fellowship programme (A. R.) and the FOCUS French LabEx doctoral fellowship program (A. R.).
E. P. acknowledges the support of the French Agence Nationale de la Recherche under grant ANR-11-BS56-015.
\end{acknowledgements}

\bibliography{biblio}

\appendix
\section{X-ray spectroscopic temperature map error estimation}\label{append:Txerror}
The X-ray spectroscopic temperature maps from \textit{Chandra} ($\TXC$) and XMM-\textit{Newton} ($\TXX$) were produced using the wavelet filtering algorithm described in \cite{Bourdin2008}. Full details of its application to the present observations can be found in \cite{Adam2016b}. As the significance of wavelet coefficients partly depends on the photon count statistics, the effective resolution varies across the map, with the higher effective area of XMM-\textit{Newton} allowing a finer sampling than \textit{Chandra} owing to its larger effective area. The pixels of the resulting maps are highly correlated because of the nature of the algorithm, which combines different scales. For this reason, we estimate the uncertainties per map pixel using a Monte Carlo approach.

In the algorithm developed by \citet{Bourdin2008}, the X-ray photons are arranged in a 3D event cube $(j,k,e)$, where $(j,k)$ are the sky coordinates and $e$ is the energy. We generated mock observation event cubes for both XMM-{\it Newton} and {\it Chandra,} where the energy coordinate $e$ of each pixel was modelled by the spectrum of the best-fitting temperature from the maps described in Sect.~\ref{sec:results}. The appropriate response function, Galactic absorption value, and redshift were folded in during this procedure. Each model spectrum was normalised to match the surface brightness in each pixel, estimated producing a wavelet cleaned, background subtracted, and exposure corrected image in the $[0.3-2.5]$ keV band.

We obtained a Monte Carlo realisation of the spectrum in each pixel to produce a new mock observation event cube. We then applied the same background subtraction procedure and wavelet filtering algorithm to this mock observation event cube, producing a new, randomised temperature map in the same way as for the real data. We did this 100 times, and took the range encompassing 68\% of the Monte Carlo realisations as the uncertainty in the temperature map.

\section{Correlation between the temperature maps}\label{append:correlation}
We performed a linear regression between the pairs of temperature maps ($\overline{T}_{1,2} \equiv \TXX, \TXC, \TSZ$), accounting for error bars on both axis. The fit is linear, but the model is not a straight line because of the zero level dependance on the effective density map (equation \ref{eq:TSZ_TX_regression}). To perform the fit, we followed \cite{Orear1982} and defined the following likelihood, $\mathscr{L}$:
\begin{equation}
2 \ {\rm ln} \ \mathscr{L} = \sum_{i=1}^{N_{\rm pix}} \frac{\left(k_{\rm B} \overline{T}^{(i)}_{1} - \alpha \ k_{\rm B} \overline{T}^{(i)}_{2} - {\beta}/{\overline{n}_{\rm e}^{(i)}}\right)^2}{\left(\delta^{(i)}_{T_{1}}\right)^2 + \left(\alpha \ \delta^{(i)}_{T_{2}}\right)^2},
\label{eq:chi2_def}
\end{equation}
where $\delta_{T_{1,2}}$ represents the temperature map uncertainties, and $\beta$ is set to zero when the regression is performed between $\TXX$ and $\TXC$. The parameter space was sampled using Markov Chains, which we evolved according to the Metropolis-Hasting algorithm \citep{Chib1995}, as carried out in \cite{Adam2015}. We checked that this method correctly reproduced the true posterior likelihood using Monte Carlo realisations (pairs of temperature maps taken as the truth, to which we added a noise realisation as expected from the error estimates). Following \citep{Pratt2009}, we computed the overall scatter as
\begin{equation}
\sigma_{\rm tot}^2 = \frac{\frac{1}{N_{\rm pix} - 2} \sum_i \frac{\left(k_{\rm B} \overline{T}^{(i)}_{1} - \alpha \ k_{\rm B} \overline{T}^{(i)}_{2} - {\beta}/{\overline{n}_{\rm e}^{(i)}}\right)^2}{\left(\delta^{(i)}_{T_{1}}\right)^2 + \left(\alpha \ \delta^{(i)}_{T_{2}}\right)^2}}{\frac{1}{N_{\rm pix}} \sum_i \frac{1}{\left(\delta^{(i)}_{T_{1}}\right)^2 + \left(\alpha \ \delta^{(i)}_{T_{2}}\right)^2}},
\label{eq:scatter_def}
\end{equation}
from which we extracted the intrinsic scatter, $\sigma_{\rm int} = \sqrt{\sigma_{\rm tot}^2 - \sigma_{\rm stat}^2}$, accounting for the statistical scatter $\sigma_{\rm stat}$.

We also checked that our posterior likelihoods were consistent with the distribution of best-fitting values obtained when fitting independently our 100 Monte Carlo map realisations ($\TX$ and $\TSZ$; see Appendix~\ref{append:Txerror} and Section~\ref{sec:Gas_mass_weighted_temperature_mapping}). Nevertheless, we stress that this fitting method does not fully account for the nature of the data themselves. Indeed the recovery of the X-ray spectroscopic temperature maps implies pixel-to-pixel correlations, which depend on the photon count statistics and thus on the sky coordinate and cluster regions. The tSZ signal is also correlated in the NIKA data, but in a different way owing to beam effects, and the noise is spatially correlated. We thus expect that the $\TX$ and $\TSZ$ map pixels do not contain the exact same sky information. These complexities, inherent to the data, are not fully accounted for in our fit. This could lead to small displacements of the best-fit values that we recover and to a slight underestimation of the error contours. However, our baseline pixel size of 20 arcsec allows us to mitigate these effects.

\end{document}

%% file: listeauthors.tex
\author{R.~Adam\inst{\ref{inst1}, \ref{inst8}}\thanks{Corresponding author: R\'emi Adam, \url{remi.adam@oca.eu}}	
\and M.~Arnaud\inst{\ref{inst2}}
\and I.~Bartalucci\inst{\ref{inst2}}
\and P.~Ade\inst{\ref{inst3}}		
\and P.~Andr\'e\inst{\ref{inst2}}		
\and A.~Beelen\inst{\ref{inst4}}		
\and A.~Beno\^it\inst{\ref{inst5}}	
\and A.~Bideaud\inst{\ref{inst3}}	
\and N.~Billot\inst{\ref{inst6}}		
\and H.~Bourdin\inst{\ref{inst7}}
\and O.~Bourrion\inst{\ref{inst8}}	
\and M.~Calvo\inst{\ref{inst5}}		
\and A.~Catalano\inst{\ref{inst8}}	
\and G.~Coiffard\inst{\ref{inst9}}	
\and B.~Comis\inst{\ref{inst8}}		
\and A.~D'Addabbo\inst{\ref{inst5}, \ref{inst10}}
\and F.-X.~D\'esert\inst{\ref{inst11}}	
\and S.~Doyle\inst{\ref{inst3}}		
\and C.~Ferrari\inst{\ref{inst1}}
\and J.~Goupy\inst{\ref{inst5}}		
\and C.~Kramer\inst{\ref{inst6}}	
\and G.~Lagache\inst{\ref{inst13}}	
\and S.~Leclercq\inst{\ref{inst9}}	
\and J.-F.~Mac\'ias-P\'erez\inst{\ref{inst8}}	
\and S.~Maurogordato\inst{\ref{inst1}}
\and P.~Mauskopf\inst{\ref{inst3}, \ref{inst14}}	
\and F.~Mayet\inst{\ref{inst8}}		
\and A.~Monfardini\inst{\ref{inst5}}	
\and F.~Pajot\inst{\ref{inst4}}		
\and E.~Pascale\inst{\ref{inst3}}	
\and L.~Perotto\inst{\ref{inst8}}		
\and G.~Pisano\inst{\ref{inst3}}		
\and E.~Pointecouteau\inst{\ref{inst16}, \ref{inst17}}
\and N.~Ponthieu\inst{\ref{inst11}}	
\and G.W.~Pratt\inst{\ref{inst2}}
\and V.~Rev\'eret\inst{\ref{inst2}}	
\and A.~Ritacco\inst{\ref{inst8}}	
\and L.~Rodriguez\inst{\ref{inst2}}	
\and C.~Romero\inst{\ref{inst9}}	
\and F.~Ruppin\inst{\ref{inst8}}		
\and K.~Schuster\inst{\ref{inst9}}	
\and A.~Sievers\inst{\ref{inst6}}	
\and S.~Triqueneaux\inst{\ref{inst5}}
\and C.~Tucker\inst{\ref{inst3}}		
\and R.~Zylka\inst{\ref{inst9}}}		

\institute{
Laboratoire Lagrange, Universit\'e C\^ote d'Azur, Observatoire de la C\^ote d'Azur, CNRS, Blvd de l'Observatoire, CS 34229, 06304 Nice cedex 4, France
  \label{inst1}
  \and
Laboratoire de Physique Subatomique et de Cosmologie, Universit\'e Grenoble Alpes, CNRS/IN2P3, 53, avenue des Martyrs, Grenoble, France
  \label{inst8}
  \and
Laboratoire AIM, IRFU/D\'epartement d'Astrophysique - CEA/DRF - CNRS - Universit\'{e} Paris Diderot, B\^{a}t. 709, CEA-Saclay, F-91191 Gif-sur-Yvette Cedex, France 
  \label{inst2}
  \and
Astronomy Instrumentation Group, University of Cardiff, UK
  \label{inst3}
\and
Institut d'Astrophysique Spatiale (IAS), CNRS and Universit\'e Paris Sud, Orsay, France
  \label{inst4}
\and
Institut N\'eel, CNRS and Universit\'e Grenoble Alpes, France
  \label{inst5}
\and
Institut de RadioAstronomie Millim\'etrique (IRAM), Granada, Spain
  \label{inst6}
\and
Dipartimento di Fisica, Universit\`a degli Studi di Roma 'Tor Vergata', via della Ricerca Scientifica, 1, I-00133 Roma, Italy
  \label{inst7}
  \and
Institut de RadioAstronomie Millim\'etrique (IRAM), Grenoble, France
  \label{inst9}
\and
Dipartimento di Fisica, Sapienza Universit\`a di Roma, Piazzale Aldo Moro 5, I-00185 Roma, Italy
  \label{inst10}
\and
Institut de Plan\'etologie et d'Astrophysique de Grenoble (IPAG), CNRS and Universit\'e Grenoble Alpes, France
  \label{inst11}
  \and
Aix Marseille Universit\'e, CNRS, LAM (Laboratoire d'Astrophysique de Marseille) UMR 7326, 13388, Marseille, France
  \label{inst13}
  \and
School of Earth and Space Exploration and Department of Physics, Arizona State University, Tempe, AZ 85287
  \label{inst14}
  \and
Universit\'e de Toulouse, UPS-OMP, Institut de Recherche en Astrophysique et Plan\'etologie (IRAP), Toulouse, France
  \label{inst16}
\and
CNRS, IRAP, 9 Av. colonel Roche, BP 44346, F-31028 Toulouse cedex 4, France 
  \label{inst17}
}